\documentclass[11pt]{llncs}
\usepackage{amssymb}
\usepackage{graphicx}
\usepackage{cite}
\usepackage{color}
\usepackage{textcomp}
\usepackage{url}
\usepackage{caption}
\usepackage{siunitx}

\usepackage{academicons}
\usepackage{hyperref}
\usepackage{xcolor}
\newcommand{\orcid}[1]{\href{https://orcid.org/#1}{\includegraphics[scale=0.08]{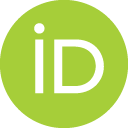}}}

\usepackage{url}
\urldef{\mailsa}\path|{alfred.hofmann, ursula.barth, ingrid.haas, frank.holzwarth,|
\urldef{\mailsb}\path|anna.kramer, leonie.kunz, christine.reiss, nicole.sator,|
\urldef{\mailsc}\path|erika.siebert-cole, peter.strasser, lncs}@springer.com|    
\newcommand{\keywords}[1]{\par\addvspace\baselineskip
\noindent\keywordname\enspace\ignorespaces#1}

\begin{document}

\mainmatter  

\title{A computational analysis of the reaction of atomic oxygen O($^3$P) with acrylonitrile}

\titlerunning{A Computational Study of the Reaction HC$_{3}$N + CN Leading to C$_{4}$N$_{2}$ + H}

\author{Luca Mancini\inst{1}\orcidID{0000-0002-9754-6071} 
\and Emília {Valença Ferreira de Aragão}\inst{1,2}\orcidID{0000-0002-8067-0914} 
}

\authorrunning{L. Mancini et al.}
%

\institute{
Dipartimento di Chimica, Biologia e Biotecnologie,\\ Universit\`{a} degli Studi di Perugia, 06123 Perugia, Italy\\ 
\email{\{emilia.dearagao,luca.mancini2\}@studenti.unipg.it}\\
\and 
Master-Tec srl, Via Sicilia 41, 06128 Perugia, Italy\\
\email{emilia.dearagao@master-tec.it}\\
}

\toctitle{Lecture Notes in Computer Science}
\tocauthor{Authors' Instructions}
\maketitle

\begin{abstract}

The work is focused on the characterization of a long-range interacting complex in the reaction between atomic oxygen, in its ground state O($^{3}$P) and acrylonitrile CH$_{2}$CHCN, also known as vinyl cyanide or cyano ethylene, through electronic structure calculations. 				
Different ab initio methods have been used in order to understand which functional provides a better description of the long-range interaction. The results of the work suggest that B2PLYPD3 gives the best description of the long-range interaction, while CAM-B3LYP represents the best compromise between chemical accuracy and computational cost.

\keywords{Ab initio calculations,  Astrochemistry, Combustion chemistry}
\end{abstract}

\section{Introduction}
The reaction between atomic oxygen, in its ground electronic state O($^3$P) and acrylonitrile can be of great interest in several fields, such as astrochemistry and combustion chemistry.

The study of planets and moons which share similarities with primitive Earth is fundamental to understand the evolution of the prebiotic chemistry in our planet, since the appearance of life has drastically changed the characteristic of Earth. One of the best candidates for this purpose is Titan, Saturn's largest moon as well as one of the few moons of the Solar System which possess a thick atmosphere.

The presence of acrylonitrile in Titan’s atmosphere has been inferred by the detection of ionic species using Cassini mass spectrometer \cite{vuitton2007ion,cui2009analysis}and later confirmed in 2017 by the first spectroscopic detection of the molecule by ALMA (Atacama Large Millimeter/submillimeter Array)\cite{palmer2017alma}. 
The reaction with atomic oxygen can be a destruction route of acrylonitrile, even though oxygenated species in the atmosphere of Titan are not abundant \cite{feuchtgruber1999oxygen,teanby2018origin,horst2008origin}. This process can be at play also in the Interstellar Medium.

The first detection of acrylonitrile goes back to 1975 towards the SgrB2 molecular cloud \cite{gardner1975detection}. This represents the first detected molecule containing a carbon-carbon double bond.
Since the first detection, the presence of acrylonitrile has been revealed in a wide variety of environments, such as the TMC-1 dark cloud where, in 1983, four rotational transitions at 3 cm and 1.5 cm have been observed for the first time \cite{matthews1983detection}.
An analysis of the excitation of several rotational transitions has been performed, in 1999, towards the SgrB2(N) hot molecular core \cite{nummelin1999vibrationally}, while later in 2008 the CH$_2$CHCN molecule has been detected in the C-rich star IRC +10216 \cite{agundez2008detection}. The analysis of the relatively high rotational temperature brought the authors to the conclusion that this specie is excited in the cyrcumstellar envelope by radiative pumping to excited vibrational states. 
Later in 2014 A. Lopez- Sepulcre et al. reported a new laboratory characterization in the 19-1983 GHz range, together with new astronomical detection between 80 and 280 GHz of acrylonitrile in its ground and vibrationally excited states with the IRAM-30m facilities \cite{lopez2014laboratory}.
A new analysis of the L1154 prestellar core has been carried out in 2019, where several N-containing species have been detected for the first time \cite{vastel2019isocyanogen}. The inventory of the aforementioned molecules includes small and simple species such ah CN and NCCN, together with more complex species like CH$_3$CN, CH$_2$CN, HCCNC and CH$_2$CHCN. A detailed chemical network has been built involving all the nitrogen bearing species detected in order to understand the isocyanogen formation in the ISM.
Moreover, the reactions of O($^3$P) with unsaturated hydrocarbons play a key role in combustion science and atmospheric chemistry \cite{cavallotti2014relevance,balucani2012crossed,cavallotti2020theoretical,leonori2012crossed,balucani2015crossed,fu2012experimental}, considering the ease with which this species is formed and the high reactivity. 
The presence of nitrogen atoms in several fuels or biomass combustion \cite{simoneit2003alkyl} makes it interesting to investigate the reaction of oxygen atoms with these substrates 

 In the present work we performed an analysis of the first step of the reaction of acrylonitrile with atomic oxygen, focusing our interest on the analysis of different theoretical methods which can be used for the identification of  a van der Waals adduct in the entrance channel.

The presence of a van der Waals adduct can affect the chemical reactivity of bimolecular recations as already find out  in several systems\cite{skouteris1999van,balucani2004dynamics,skouteris2001experimental,heard2018rapid,efficient}.

 The title reaction has been already investigated theoretically BY J. Sun et al. \cite{sun2015theoretical} who performed an exploration of the triplet potential energy surface. An experimental characterization has been performed by H. P. Upadhyay et al. \cite{upadhyaya1997reaction} in a flow discharge tube using the O($^3$P) chemiluminescence titration method.
More generally it is reasonable to think that the formation of a van der Waals adduct is one of the first stages of most of the reactions. The long-range interaction in the  formed complex can lead to molecular geometries that can promote or hinder the evolution of the reaction. As a consequence the formation of a van der Waals complex can strongly affect the rate constants. Unfortunately the identification of the aforementioned complexes appears to be difficult with  mostly used ab initio methods. In the following sections a comparison between the results obtained at different levels of theory is presented. 
In particular the analysis started with a benchmark work in order to compare different methods, see next sections for more details, while the last part of the work is focused on the comparison between two particular level of theory: B3LYP and CAM-B3LYP.

\section{Methods}
The investigation of the title system has been performed adopting a computational strategy which has been successfully used in several cases \cite{falcinelli2016stereoselectivity,leonori2009crossed,bartolomei2008intermolecular,de2011proton,leonori2009observation,de2007ssoh,rosi2012theoretical,berteloite2011low,rosi2013theoretical,sleiman2018low}. 
In particular electronic structure calculations have been performed for the reactants and for the long-range complex on the overall triplet Potential Energy Surface (PES). 
In all calculations, the geometries of the stationary points were treated with two different methods: one for optimization and another to obtain more accurate energy values.
Geometry optimizations were performed in order to benchmark several methods: density functional theory (DFT), with the Becke-3-parameter exchange and Lee-Yang- Parr correlation (B3LYP) \cite{becke1993density,stephens1994ab} combined or not the with Grimme’s D3BJ \cite{grimme2011effect,goerigk2011efficient} dispersion (B3LYPD3); Coulomb Attenuating Method (CAM-B3LYP) \cite{yanai2004new}; double-hybrid DFT method B2PLYP \cite{grimme2006semiempirical} combined or not with Grimme’s D3BJ dispersion (B2PLYPD3) and the long-range corrected functional wB97X \cite{chai2008systematic} also with the inclusion of a version of Grimme's D2 dispersion model (wB97XD) \cite{chai2008long}.
In particular the aforementioned functional, named CAM-B3LYP (Coulomb Attenuating Method-B3LYP) represents a new hybrid exchange-correlation functional with improved long-range properties with respect to the B3LYP \cite{yanai2004new}.
In details, the electron repulsion operator $\frac{1}{r_{12}}$ , which was already divided into short-range and long-range parts by Tsuneda and collaborators~\cite{tawada2004long} as follows:

\begin{equation}
\frac{1}{r_{12}}=\frac{1-erf({\mu}r_{12})}{r_{12}}+\frac{erf({\mu}r_{12})}{r_{12}}
\label{eq:1}
\end{equation}
\noindent

is now implemented using two parameters $\alpha$ and $\beta$:

\begin{equation}
\frac{1}{r_{12}}=\frac{1-[\alpha+\beta*erf({\mu}r_{12})]}{r_{12}}+\frac{\alpha+\beta*erf({\mu}r_{12})}{r_{12}}
\label{eq:2}
\end{equation}
\noindent

where 0$\leq$$\alpha$+$\beta$$\leq$1 together with 0$\leq$$\alpha$$\leq$1 and 0$\leq$$\beta$$\leq$1.

All the above mentioned methods have been used in conjunction with the correlation consistent valence polarized basis set aug-cc-pVTZ \cite{dunning1989gaussian}. The same level of theory used for geometry optimization was used to perform an harmonic vibrational frequency analysis in order to assign the nature of each identified stationary point (i.e. minimum if all the frequencies are real and saddle point if there is one, and only one, imaginary frequency). Then for each stationary point for all the employed methods, a more accurate energy values was computed at coupled cluster level, including single and double excitations as well as perturbative estimate of connected triples (CCSD(T))\cite{bartlett1981many,raghavachari1989fifth,olsen1996full}. 
Finally, energies obtained from every method were corrected to 0 K by adding the zero-point energy correction that had been derived from the frequency calculations.
All the calculations were performed using the Gaussian09 software\cite{frisch2009gaussian}, while the analysis of the vibrational frequencies was done using Avogadro\cite{hanwell2012avogadro}.

\begin{figure}[h]
\centering
\includegraphics[scale=0.3]{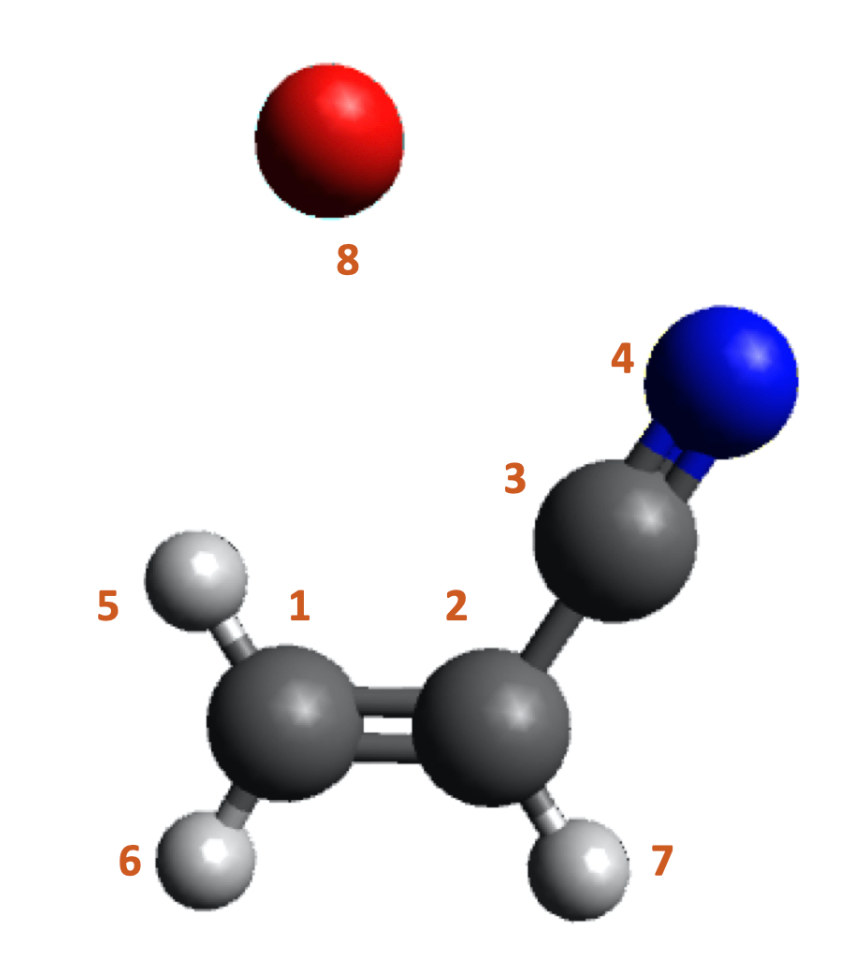}
\caption{Geometry of the long-range identified complex}
\label {fig1}
\end{figure}

\section{Results and discussion}

The data obtained from electronic structure calculations clearly shows the formation of a long-range complex as first step of the reaction between O($^3$P) and acrylonitrile. The process appears to be exothermic at all the previously cited levels of theory. The structure of the complex shows the interaction of the oxygen atom with the double bond between C$_{1}$ and C$_{2}$ of the CH$_2$CHCN molecule. In tab 1 are reported the bond distances, expressed in angstroms, obtained at the different level of calculations. 
The geometry of the complex with all the atom labels is reported in fig.1. A comparison between the values of bond distances obtained with the seven different theoretical methods can bring to the conclusion that there are no significant differences in the structure of the van der Waals complex at all the levels of calculation. The main discrepancies can be observed in the values of distance related to the long-range interaction between the oxygen atom and the three C atom of the molecule where, however, the largest deviation is around 0.2 Å. 
In order to have a better comprehension of the differences between the methods is possible to compare the energies obtained from the various calculations.

\begin{table}[t]
\centering
\caption{Bond distances, in Å, obtained at the different level of calculations}
\label {Tab.1}
\includegraphics[scale=0.35]{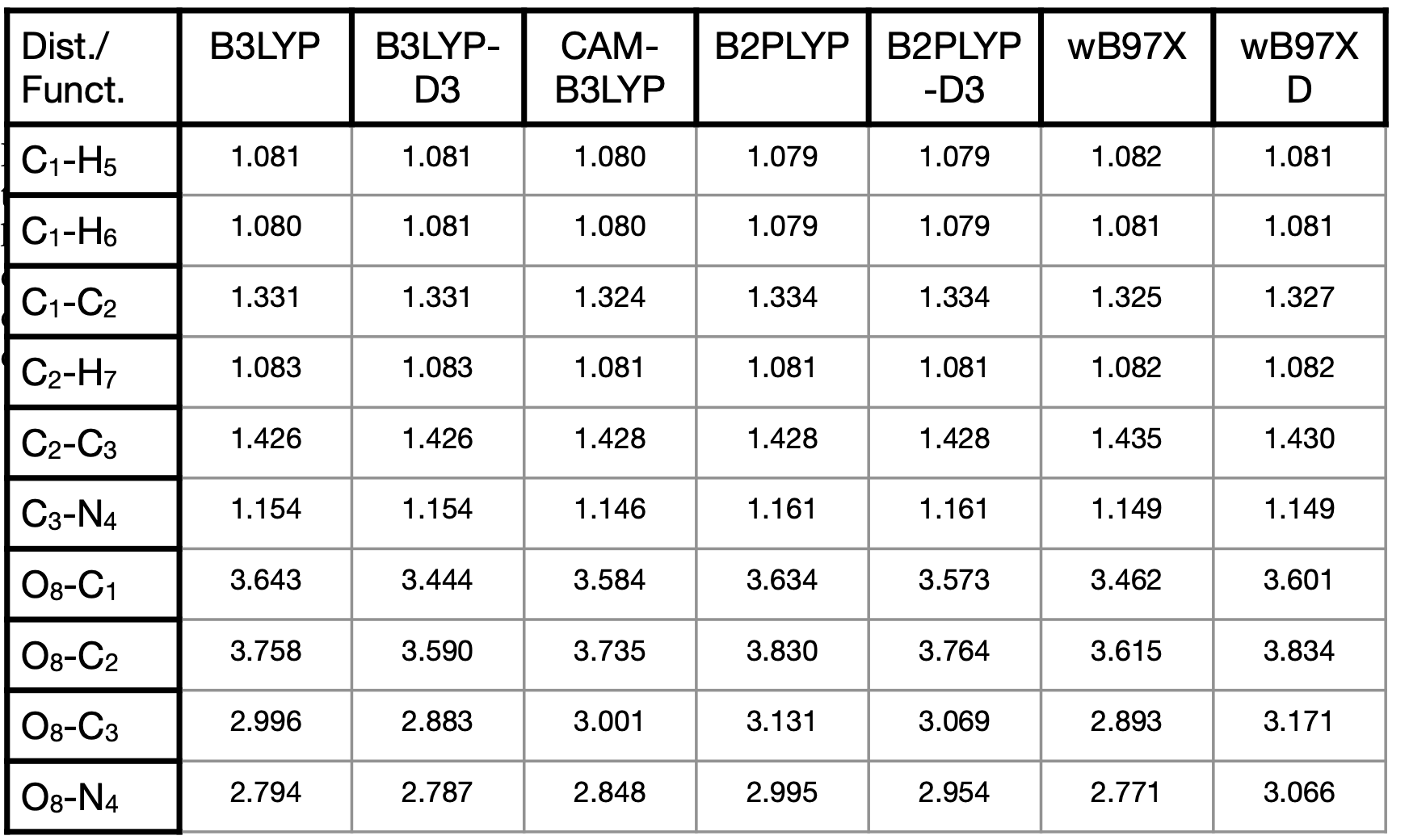}
\end{table}

In tab. 2 the energies obtained at the different levels of theory are reported. 
In particular the first column reports the electronic energies computed at the defined level of theory (in kJ/mol), while in the last  column are reported the values obtained by the CCSD(T) calculations starting from the geometry optimized at the lower level of theory. 
The values are corrected at 0 K including the zero point correction obtained  from the harmonic vibrational calculations performed at the same level of theory used for the geometry optimization.
\begin{table}[t]
\centering
\caption{Energies obtained at the different level of calculations.}
\label {Tab.2}
\includegraphics[scale=0.35]{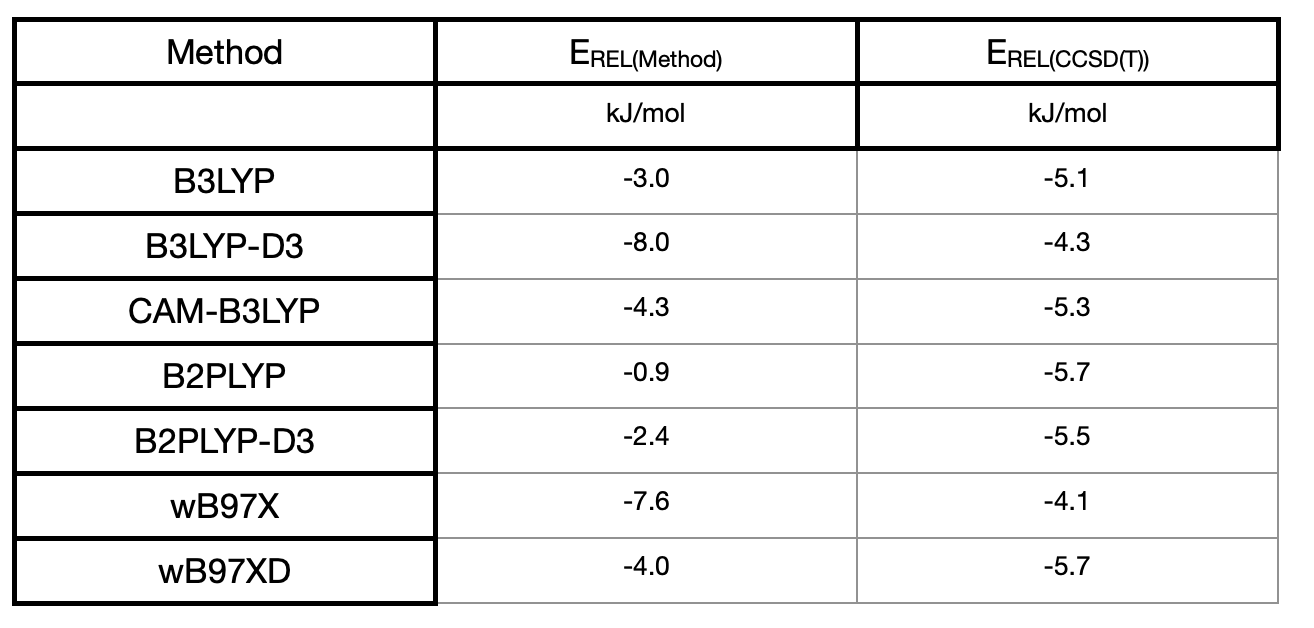}
\end{table}
The analysis of the values of energy reported in tab. 2 shows no significant differences in the values of the CCSD(T) corrected energy, which is usually considered in the construction of the potential energy surface. 
In particular we can notice that most of the time the difference is lower than 5 kJ/mol, which is considered to be the uncertainty associated to accurate calculations. 

Since no significant differences can be noticed between the employed methods we decided to focus our attention on the comparison between the B3LYP method and CAM-B3LYP, which is presented as an improvement of the B3LYP in order to include the long-range interaction. 
In fig 2 and 3 a comparison of the bond distances (in Å) and angles (in °) between the two methods is reported for the reactant (vinyl cyanide) and for the complex respectively. 
The results of the B3LYP/aug-cc-pVTZ analysis are shown in blue while the values derived from the CAM-B3LYP/aug-cc-pVTZ calculations are displayed in green.

\begin{figure}[t]
\centering
\includegraphics[scale=0.35]{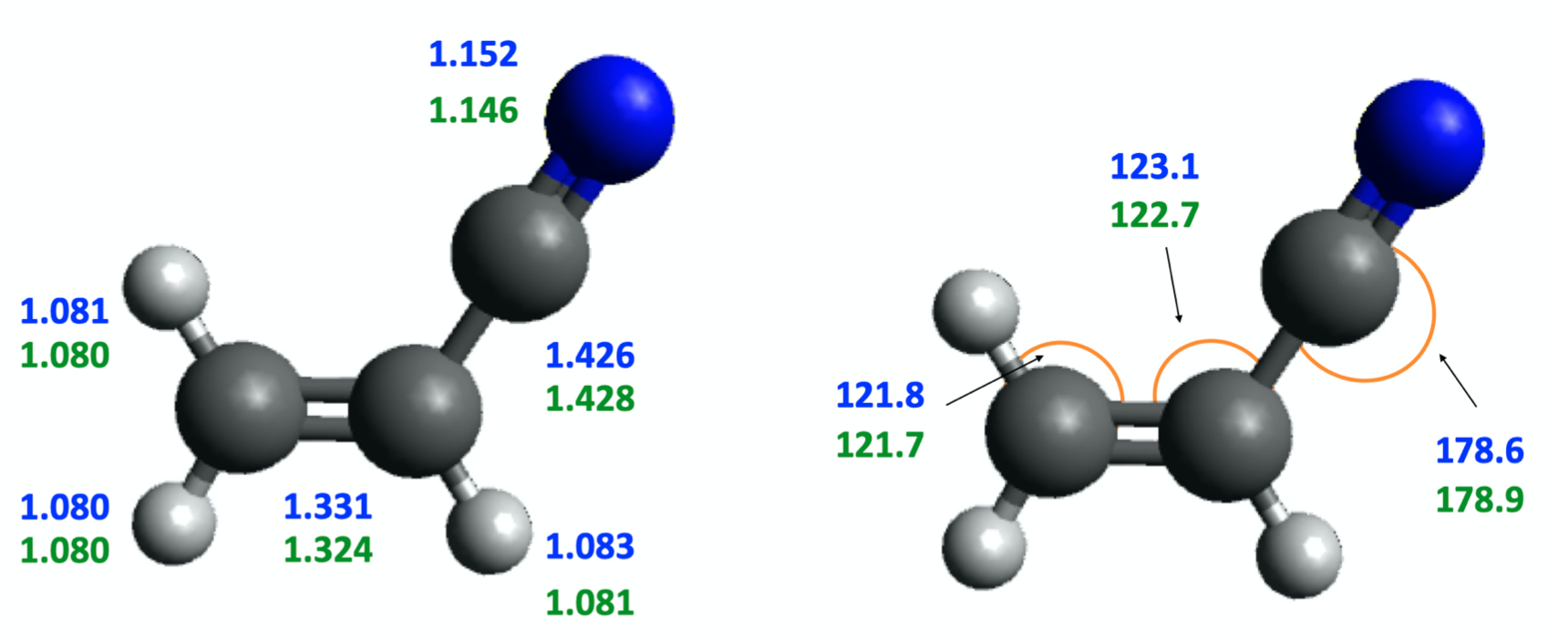}
\caption{Bond lengths (Å) and angles for the reactant at the B3LYP/aug-cc-pVTZ (blue) and CAM- B3LYP/aug-cc-pVTZ (green) level of theory.}
\label {fig2}
\end{figure}

\begin{figure}[t]
\centering
\includegraphics[scale=0.35]{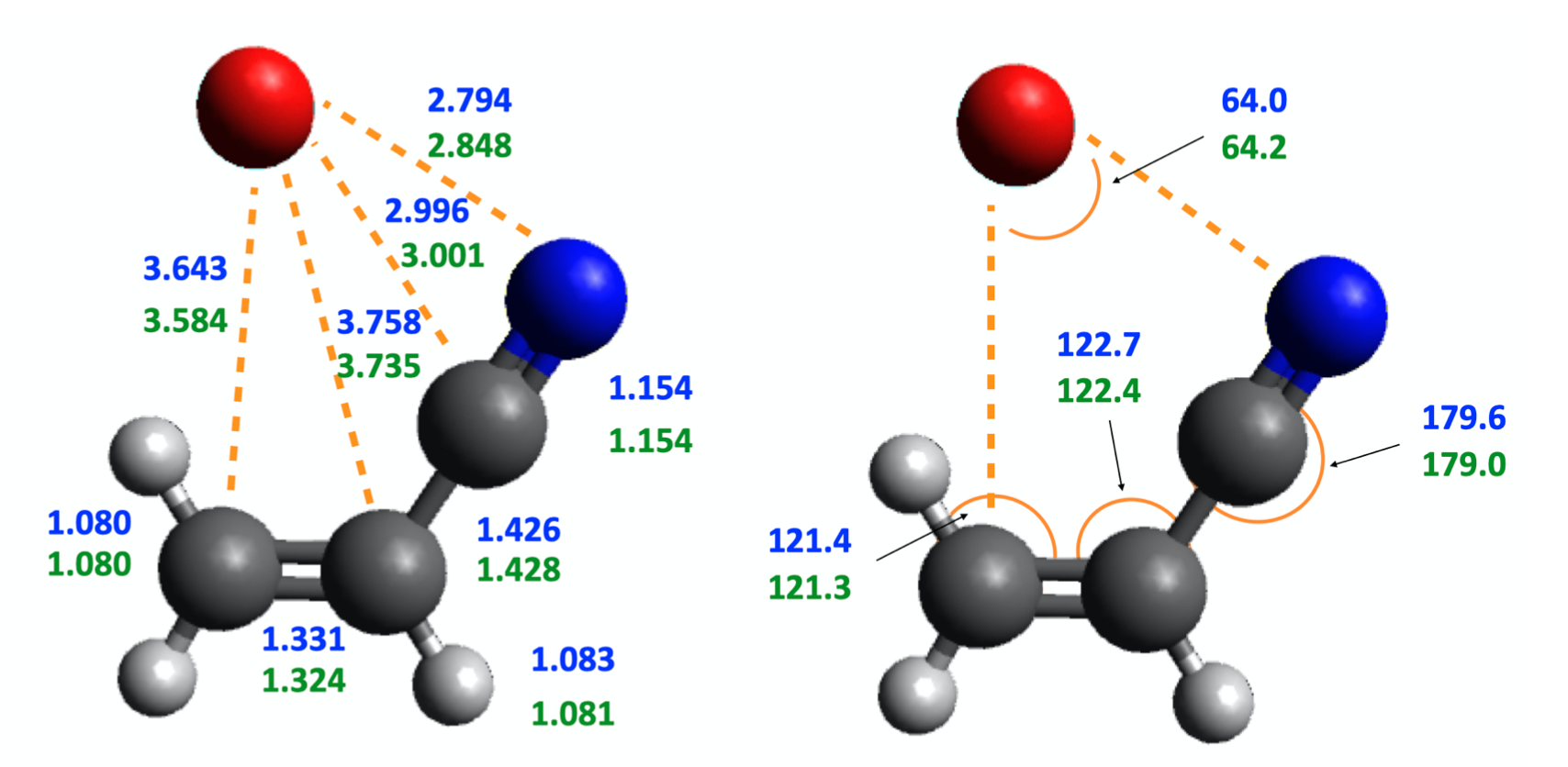}
\caption{Bond lengths (Å) and angles for the vdW adduct at the B3LYP/aug-cc-pVTZ (blue) and CAM-B3LYP/aug-cc-pVTZ (green) level of theory.}
\label {fig3}
\end{figure}

Also in this case the main differences can be appreciated in the description of the long-range O-C interaction with deviations lower than 0.1 Å for the bond distances and of a maximum of 0.6° for the angles. 
The analysis of the harmonic vibrational frequencies, reported in tab.3, shows small differences between the two levels of theory. These differences lead to a small variation on the zero-point correction which is equal to 0.051064 (Hartree/Particle) at B3LYP/aug-cc-pVTZ and 0.051720 (Hartree/Particle) at CAM-B3LYP/aug-cc-pVTZ level of theory.

\begin{table}[h]
\centering
\caption{Harmonic vibrational frequencies (cm-1) of the vdW adduct obtained at the B3LYP/aug-cc-pVTZ and CAM-B3LYP/aug-cc-pVTZ level of theory.}
\label {Tab.3}
\includegraphics[scale=0.5]{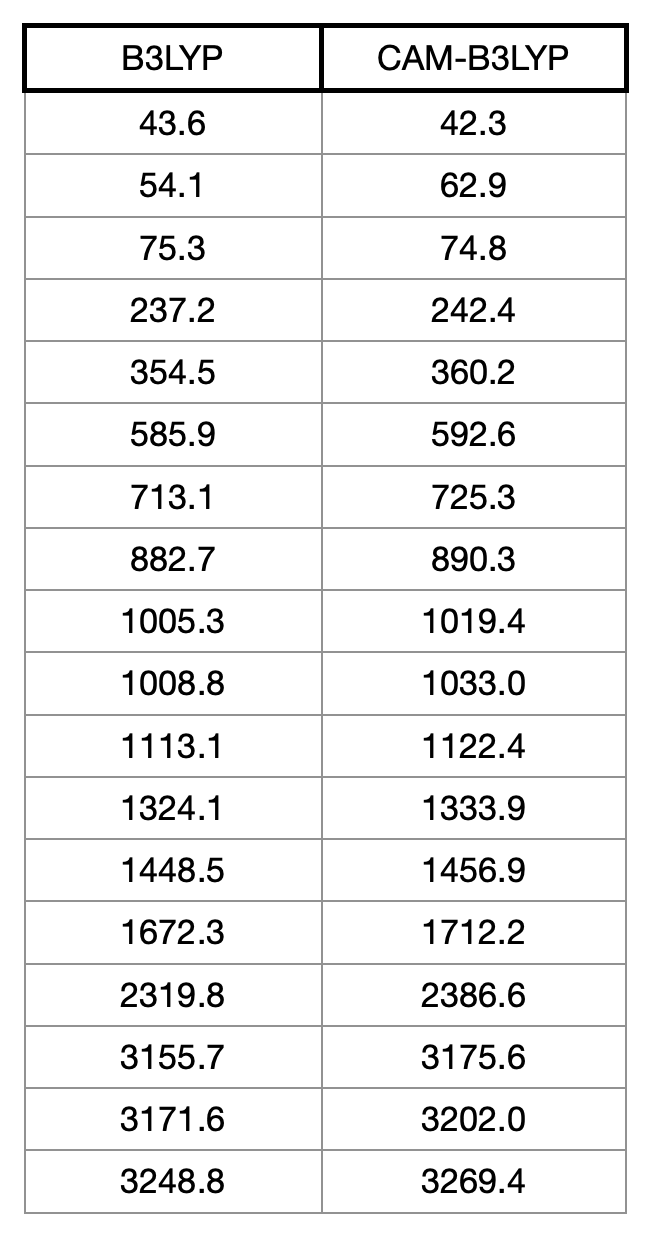}
\end{table}

\begin{figure}[h]
\centering
\includegraphics[scale=0.26]{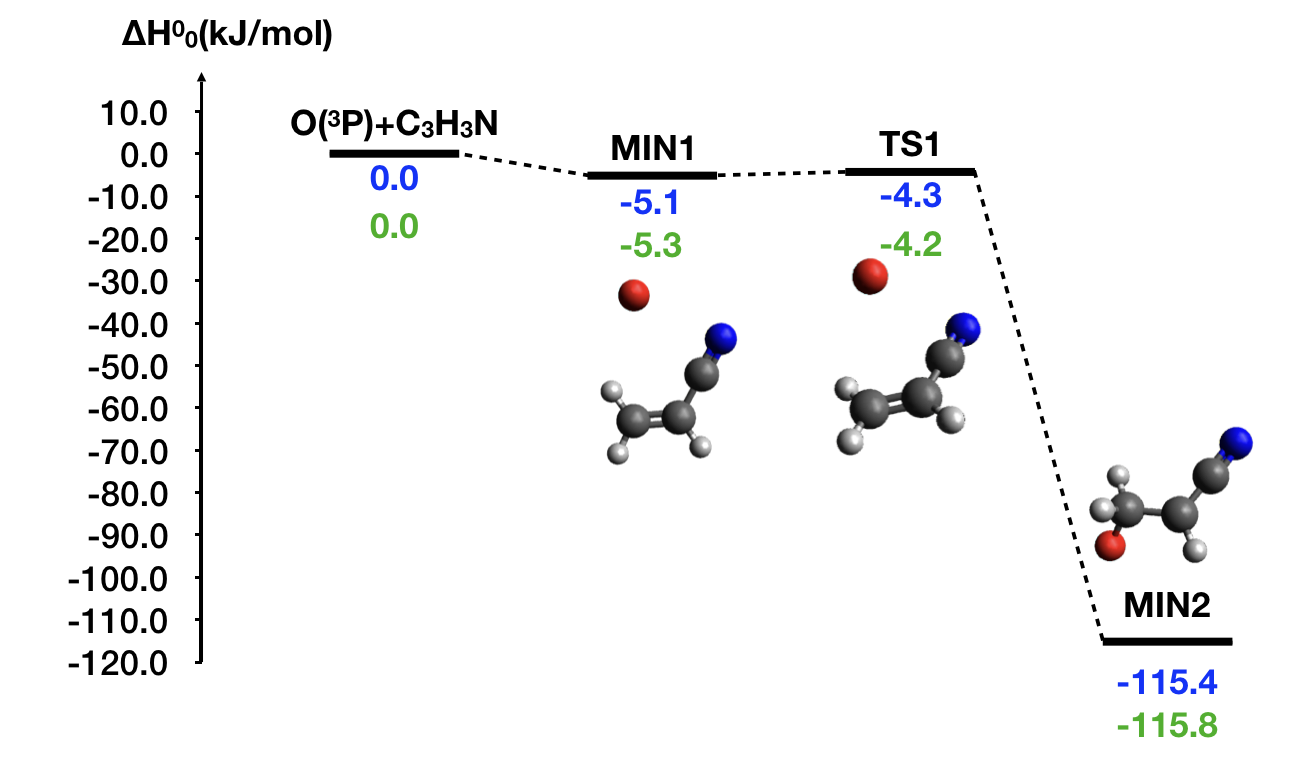}
\caption{Schematic representation of the potential energy surface for the first steps of the attack of the O($^{3}$P) atom to the acrylonitrile molecule. Energies computed at CCSD(T)/aug-cc-pVTZ level with zero-point corrections at B3LYP/aug-cc-pVTZ (blue) and CAM-B3LYP/aug-cc-pVTZ (green) levels. }
\label {fig4}
\end{figure}

A last comparison between the two methods can be performed considering the first steps of the reaction, shown in fig. 4. 
The reaction starts with a barrierless formation of the previously described complex, MIN1, followed by the formation of the minimum MIN2 in which we can notice the formation of a chemical bond between C$_{1}$ and O, through a small barrier, represented by TS1. 
No particular differences can be noticed in the values of energy obtained at CCSD(T)/aug-cc-pVTZ level considering the geometries optimized at B3LYP level (in blue) and at CAM-B3LYP (in green) level of theory.

\section{Conclusions}

In the present work we performed a benchmark analysis of the first step of the reaction between atomic oxygen O($^3$P) and acrylonitrile (CH$_2$CHCN) using different theoretical methods: DFT (with B3LYP and B3LYPD3 functional), double-hybrid DFT (with B2PLYP and B2PLYPD3 functionals) and the long-range corrected functional wB97X and wB97XD. 
As far as optimized geometries for stationary points are concerned, all the methods seem to  provide similar results. In particular, B3LYP functional appears to be a good compromise between accuracy and computational cost for the characterization of minima. 
The analysis of the long-range interaction can be performed also using the CAM-B3LYP functionals, which provides better estimate of the interactions than the B3LYP functional but needs lower computational resources than the B2PLYPD3 functional. The best choice for the estimate of the energies appears to be the use of correlated methods like CCSD(T). 
A more general conclusion can be presented concerning the first steps of the reaction mechanism, which appears to be in agreement with the previous determination by J. Sun et al. \cite{sun2015theoretical}. In particular the oxygen atom starts to interact with the electron density of multiple bonds in the molecule to form an exothermic addition intermediate. The comparison with B3LYP and CAM-B3LYP functionals suggests the possibility to use a combined strategy for the analysis of the Potential Energy Surfaces which will allow us to locate the long-range complexes at CAM-B3LYP level to be included in the reaction pathways.

\section{Acknowledgements}

This project has received funding from the European Union’s Horizon 2020 research and innovation programme under the Marie Skłodowska-Curie grant agreement No 811312 for the project "Astro-Chemical Origins” (ACO).
 This work was supported by the Italian Space Agency (ASI) BANDO ASI DC-VUM-2017-034 CONTRATTO DI FINANZIAMENTO ASI N. 2019-3 U.O, CUP F86C16000000006  "Vita nello spazio - Origine, presenza, persistenza della vita nello spazio, dalle molecole agli estremofili"
 The authors acknowledge the Dipartimento di Ingegneria Civile e Ambientale of the University of Perugia
for allocated computing time within the project “Dipartimenti di Eccellenza 2018-2022”.

\bibliographystyle{splncs04} 
\bibliography{Acrylonitrile_oxygen_ICCSA2021}{}

\begin{thebibliography}{10}
\providecommand{\url}[1]{\texttt{#1}}
\providecommand{\urlprefix}{URL }
\providecommand{\doi}[1]{https://doi.org/#1}

\bibitem{vuitton2007ion}
Vuitton, V., Yelle, R., McEwan, M.: Ion chemistry and n-containing molecules in
  titan's upper atmosphere. Icarus  \textbf{191}(2),  722--742 (2007)

\bibitem{cui2009analysis}
Cui, J., Yelle, R., Vuitton, V., Waite~Jr, J., Kasprzak, W., Gell, D., Niemann,
  H., M{\"u}ller-Wodarg, I., Borggren, N., Fletcher, G., et~al.: Analysis of
  titan's neutral upper atmosphere from cassini ion neutral mass spectrometer
  measurements. Icarus  \textbf{200}(2),  581--615 (2009)

\bibitem{palmer2017alma}
Palmer, M.Y., Cordiner, M.A., Nixon, C.A., Charnley, S.B., Teanby, N.A.,
  Kisiel, Z., Irwin, P.G., Mumma, M.J.: Alma detection and astrobiological
  potential of vinyl cyanide on titan. Science advances  \textbf{3}(7),
  e1700022 (2017)

\bibitem{feuchtgruber1999oxygen}
Feuchtgruber, H., Lellouch, E., Encrenaz, T., B{\'e}zard, B., Coustenis, A.,
  Drossart, P., Salama, A., De~Graauw, T., Davis, G.: Oxygen in the
  stratospheres of the giant planets and titan. In: The Universe as seen by
  ISO. vol.~427, p.~133 (1999)

\bibitem{teanby2018origin}
Teanby, N., Cordiner, M.A., Nixon, C.A., Irwin, P., H{\"o}rst, S., Sylvestre,
  M., Serigano, J., Thelen, A., Richards, A., Charnley, S.: The origin of
  titan’s external oxygen: further constraints from alma upper limits on cs
  and ch2nh. The Astronomical Journal  \textbf{155}(6), ~251 (2018)

\bibitem{horst2008origin}
H{\"o}rst, S.M., Vuitton, V., Yelle, R.V.: Origin of oxygen species in titan's
  atmosphere. Journal of Geophysical Research: Planets  \textbf{113}(E10)
  (2008)

\bibitem{gardner1975detection}
Gardner, F., Winnewisser, G.: The detection of interstellar vinyl
  cyanide/acrylonitrile. The Astrophysical Journal  \textbf{195},  L127--L130
  (1975)

\bibitem{matthews1983detection}
Matthews, H.E., Sears, T.J.: The detection of vinyl cyanide in tmc-1. The
  Astrophysical Journal  \textbf{272},  149--153 (1983)

\bibitem{nummelin1999vibrationally}
Nummelin, A., Bergman, P.: Vibrationally excited vinyl cyanide in sgr b2 (n).
  Astronomy and Astrophysics  \textbf{341},  L59--L62 (1999)

\bibitem{agundez2008detection}
Ag{\'u}ndez, M., Fonfr{\'\i}a, J.P., Cernicharo, J., Pardo, J., Gu{\'e}lin, M.:
  Detection of circumstellar ch2chcn, ch2cn, ch3cch, and h2cs. Astronomy \&
  Astrophysics  \textbf{479}(2),  493--501 (2008)

\bibitem{lopez2014laboratory}
L{\'o}pez, A., Tercero, B., Kisiel, Z., Daly, A.M., Berm{\'u}dez, C., Calcutt,
  H., Marcelino, N., Viti, S., Drouin, B., Medvedev, I., et~al.: Laboratory
  characterization and astrophysical detection of vibrationally excited states
  of vinyl cyanide in orion-kl. Astronomy \& Astrophysics  \textbf{572}, ~A44
  (2014)

\bibitem{vastel2019isocyanogen}
Vastel, C., Loison, J.C., Wakelam, V., Lefloch, B.: Isocyanogen formation in
  the cold interstellar medium. Astronomy \& Astrophysics  \textbf{625}, ~A91
  (2019)

\bibitem{cavallotti2014relevance}
Cavallotti, C., Leonori, F., Balucani, N., Nevrly, V., Bergeat, A., Falcinelli,
  S., Vanuzzo, G., Casavecchia, P.: Relevance of the channel leading to
  formaldehyde+ triplet ethylidene in the o (3p)+ propene reaction under
  combustion conditions. The journal of physical chemistry letters
  \textbf{5}(23),  4213--4218 (2014)

\bibitem{balucani2012crossed}
Balucani, N., Leonori, F., Casavecchia, P.: Crossed molecular beam studies of
  bimolecular reactions of relevance in combustion. Energy  \textbf{43}(1),
  47--54 (2012)

\bibitem{cavallotti2020theoretical}
Cavallotti, C., De~Falco, C., Pratali~Maffei, L., Caracciolo, A., Vanuzzo, G.,
  Balucani, N., Casavecchia, P.: Theoretical study of the extent of intersystem
  crossing in the o (3p)+ c6h6 reaction with experimental validation. The
  Journal of Physical Chemistry Letters  \textbf{11}(22),  9621--9628 (2020)

\bibitem{leonori2012crossed}
Leonori, F., Occhiogrosso, A., Balucani, N., Bucci, A., Petrucci, R.,
  Casavecchia, P.: Crossed molecular beam dynamics studies of the o (3p)+
  allene reaction: primary products, branching ratios, and dominant role of
  intersystem crossing. The Journal of Physical Chemistry Letters
  \textbf{3}(1),  75--80 (2012)

\bibitem{balucani2015crossed}
Balucani, N., Leonori, F., Casavecchia, P., Fu, B., Bowman, J.M.: Crossed
  molecular beams and quasiclassical trajectory surface hopping studies of the
  multichannel nonadiabatic o (3p)+ ethylene reaction at high collision energy.
  The Journal of Physical Chemistry A  \textbf{119}(50),  12498--12511 (2015)

\bibitem{fu2012experimental}
Fu, B., Han, Y.C., Bowman, J.M., Leonori, F., Balucani, N., Angelucci, L.,
  Occhiogrosso, A., Petrucci, R., Casavecchia, P.: Experimental and theoretical
  studies of the o (3p)+ c2h4 reaction dynamics: Collision energy dependence of
  branching ratios and extent of intersystem crossing. The Journal of chemical
  physics  \textbf{137}(22),  22A532 (2012)

\bibitem{simoneit2003alkyl}
Simoneit, B.R., Rushdi, A., Bin~Abas, M., Didyk, B.: Alkyl amides and nitriles
  as novel tracers for biomass burning. Environmental science \& technology
  \textbf{37}(1),  16--21 (2003)

\bibitem{skouteris1999van}
Skouteris, D., Manolopoulos, D.E., Bian, W., Werner, H.J., Lai, L.H., Liu, K.:
  van der waals interactions in the cl+ hd reaction. Science
  \textbf{286}(5445),  1713--1716 (1999)

\bibitem{balucani2004dynamics}
Balucani, N., Skouteris, D., Capozza, G., Segoloni, E., Casavecchia, P.,
  Alexander, M.H., Capecchi, G., Werner, H.J.: The dynamics of the prototype
  abstraction reaction cl (2 p 3/2, 1/2)+ h 2: A comparison of crossed
  molecular beam experiments with exact quantum scattering calculations on
  coupled ab initio potential energy surfaces. Physical Chemistry Chemical
  Physics  \textbf{6}(21),  5007--5017 (2004)

\bibitem{skouteris2001experimental}
Skouteris, D., Werner, H.J., Aoiz, F.J., Banares, L., Castillo, J.F.,
  Men{\'e}ndez, M., Balucani, N., Cartechini, L., Casavecchia, P.: Experimental
  and theoretical differential cross sections for the reactions cl+ h 2/d 2.
  The Journal of Chemical Physics  \textbf{114}(24),  10662--10672 (2001)

\bibitem{heard2018rapid}
Heard, D.E.: Rapid acceleration of hydrogen atom abstraction reactions of oh at
  very low temperatures through weakly bound complexes and tunneling. Accounts
  of chemical research  \textbf{51}(11),  2620--2627 (2018)

\bibitem{efficient}
Recio,P., Alessandrini,S., Marchione,D.,Caracciolo,A., Murray,v.J.,
  Casavecchia,P, Balucani,N., Baggioli,A., Cavallotti,C., Puzzarini,C.,
  Barone,V.: Efficient Intersystem Crossing from Weakly Bound pre-Reactive
  Complex Avoids the Entrance Barrier of Bimolecular Reactions. Nature
  Chemistry (submitted)

\bibitem{sun2015theoretical}
Sun, J., Wu, W., Zhang, Y., Tang, Y., Yi, H., Wang, R., et~al.: Theoretical
  investigation on atmospheric reaction of atomic o (3p) with acrylonitrile.
  Computational and Theoretical Chemistry  \textbf{1052},  17--25 (2015)

\bibitem{upadhyaya1997reaction}
Upadhyaya, H.P., Naik, P.D., Pavanaja, U.B., Kumar, A., Vatsa, R.K., Sapre,
  A.V., Mittal, J.P.: Reaction kinetics of o (3p) with acrylonitrile and
  crotononitrile. Chemical physics letters  \textbf{274}(4),  383--389 (1997)

\bibitem{falcinelli2016stereoselectivity}
Falcinelli, S., Rosi, M., Cavalli, S., Pirani, F., Vecchiocattivi, F.:
  Stereoselectivity in autoionization reactions of hydrogenated molecules by
  metastable noble gas atoms: the role of electronic couplings. Chemistry--A
  European Journal  \textbf{22}(35),  12518--12526 (2016)

\bibitem{leonori2009crossed}
Leonori, F., Petrucci, R., Balucani, N., Hickson, K.M., Hamberg, M., Geppert,
  W.D., Casavecchia, P., Rosi, M.: Crossed-beam and theoretical studies of the
  {S($^{1}$D) + C$_{2}$H$_{2}$} reaction. The Journal of Physical Chemistry A
  \textbf{113}(16),  4330--4339 (2009)

\bibitem{bartolomei2008intermolecular}
Bartolomei, M., Cappelletti, D., de~Petris, G., Teixidor, M.M., Pirani, F.,
  Rosi, M., Vecchiocattivi, F.: The intermolecular potential in {NO-N$_{2}$ and
  (NO-N$_{2}$)$^{+}$} systems: implications for the neutralization of ionic
  molecular aggregates. Physical Chemistry Chemical Physics  \textbf{10}(39),
  5993--6001 (2008)

\bibitem{de2011proton}
de~Petris, G., Cartoni, A., Rosi, M., Barone, V., Puzzarini, C., Troiani, A.:
  {The Proton Affinity and Gas-Phase Basicity of Sulfur Dioxide}. ChemPhysChem
  \textbf{12}(1),  112--115 (2011)

\bibitem{leonori2009observation}
Leonori, F., Petrucci, R., Balucani, N., Casavecchia, P., Rosi, M., Berteloite,
  C., Le~Picard, S.D., Canosa, A., Sims, I.R.: Observation of organosulfur
  products (thiovinoxy, thioketene and thioformyl) in crossed-beam experiments
  and low temperature rate coefficients for the reaction {S($^{1}$D) +
  C$_{2}$H$_{4}$}. Physical Chemistry Chemical Physics  \textbf{11}(23),
  4701--4706 (2009)

\bibitem{de2007ssoh}
de~Petris, G., Rosi, M., Troiani, A.: {SSOH and HSSO Radicals: An Experimental
  and Theoretical Study of [S$_{2}$OH]$^{0/+/-}$ Species}. The Journal of
  Physical Chemistry A  \textbf{111}(28),  6526--6533 (2007)

\bibitem{rosi2012theoretical}
Rosi, M., Falcinelli, S., Balucani, N., Casavecchia, P., Leonori, F.,
  Skouteris, D.: {Theoretical Study of Reactions Relevant for Atmospheric
  Models of Titan: Interaction of Excited Nitrogen Atoms with Small
  Hydrocarbons}. In: Murgante, B., Gervasi, O., Misra, S., Nedjah, N., Rocha,
  A.M.A.C., Taniar, D., Apduhan, B.O. (eds.) Computational Science and Its
  Applications -- ICCSA 2012. pp. 331--344. Springer Berlin Heidelberg, Berlin,
  Heidelberg (2012)

\bibitem{berteloite2011low}
Berteloite, C., Le~Picard, S.D., Sims, I.R., Rosi, M., Leonori, F., Petrucci,
  R., Balucani, N., Wang, X., Casavecchia, P.: Low temperature kinetics,
  crossed beam dynamics and theoretical studies of the reaction s (1 d)+ ch 4
  and low temperature kinetics of s (1 d)+ c 2 h 2. Physical Chemistry Chemical
  Physics  \textbf{13}(18),  8485--8501 (2011)

\bibitem{rosi2013theoretical}
Rosi, M., Falcinelli, S., Balucani, N., Casavecchia, P., Skouteris, D.: A
  theoretical study of formation routes and dimerization of methanimine and
  implications for the aerosols formation in the upper atmosphere of titan. In:
  International Conference on Computational Science and Its Applications. pp.
  47--56. Springer (2013)

\bibitem{sleiman2018low}
Sleiman, C., El~Dib, G., Rosi, M., Skouteris, D., Balucani, N., Canosa, A.: Low
  temperature kinetics and theoretical studies of the reaction cn+ ch 3 nh 2: a
  potential source of cyanamide and methyl cyanamide in the interstellar
  medium. Physical Chemistry Chemical Physics  \textbf{20}(8),  5478--5489
  (2018)

\bibitem{becke1993density}
Becke, A.D.: Density functional thermochemistry. {III. The} role of exact
  exchange. The Journal of Chemical Physics  \textbf{98}(7),  5648--5652
  (1993). \doi{10.1063/1.464913}, \url{https://doi.org/10.1063/1.464913}

\bibitem{stephens1994ab}
Stephens, P.J., Devlin, F.J., Chabalowski, C.F., Frisch, M.J.: {${Ab}$
  ${Initio}$ Calculation of Vibrational Absorption and Circular Dichroism
  Spectra Using Density Functional Force Fields}. The Journal of physical
  chemistry  \textbf{98}(45),  11623--11627 (1994)

\bibitem{grimme2011effect}
Grimme, S., Ehrlich, S., Goerigk, L.: Effect of the damping function in
  dispersion corrected density functional theory. Journal of computational
  chemistry  \textbf{32}(7),  1456--1465 (2011)

\bibitem{goerigk2011efficient}
Goerigk, L., Grimme, S.: {Efficient and Accurate Double-Hybrid-Meta-GGA Density
  Functionals- Evaluation with the Extended GMTKN30 Database for General Main
  Group Thermochemistry, Kinetics, and Noncovalent Interactions}. Journal of
  chemical theory and computation  \textbf{7}(2),  291--309 (2011)

\bibitem{yanai2004new}
Yanai, T., Tew, D.P., Handy, N.C.: A new hybrid exchange--correlation
  functional using the coulomb-attenuating method (cam-b3lyp). Chemical physics
  letters  \textbf{393}(1-3),  51--57 (2004)

\bibitem{grimme2006semiempirical}
Grimme, S.: Semiempirical hybrid density functional with perturbative
  second-order correlation. The Journal of chemical physics  \textbf{124}(3),
  034108 (2006)

\bibitem{chai2008systematic}
Chai, J.D., Head-Gordon, M.: Systematic optimization of long-range corrected
  hybrid density functionals. The Journal of chemical physics  \textbf{128}(8),
   084106 (2008)

\bibitem{chai2008long}
Chai, J.D., Head-Gordon, M.: Long-range corrected hybrid density functionals
  with damped atom--atom dispersion corrections. Physical Chemistry Chemical
  Physics  \textbf{10}(44),  6615--6620 (2008)

\bibitem{tawada2004long}
Tawada, Y., Tsuneda, T., Yanagisawa, S., Yanai, T., Hirao, K.: A
  long-range-corrected time-dependent density functional theory. The Journal of
  chemical physics  \textbf{120}(18),  8425--8433 (2004)

\bibitem{dunning1989gaussian}
Dunning~Jr, T.H.: Gaussian basis sets for use in correlated molecular
  calculations. {I.} the atoms boron through neon and hydrogen. The Journal of
  chemical physics  \textbf{90}(2),  1007--1023 (1989)

\bibitem{bartlett1981many}
Bartlett, R.J.: Many-body perturbation theory and coupled cluster theory for
  electron correlation in molecules. Annual Review of Physical Chemistry
  \textbf{32}(1),  359--401 (1981)

\bibitem{raghavachari1989fifth}
Raghavachari, K., Trucks, G.W., Pople, J.A., Head-Gordon, M.: A fifth-order
  perturbation comparison of electron correlation theories. Chemical Physics
  Letters  \textbf{157}(6),  479--483 (1989)

\bibitem{olsen1996full}
Olsen, J., J{\o}rgensen, P., Koch, H., Balkova, A., Bartlett, R.J.: Full
  configuration--interaction and state of the art correlation calculations on
  water in a valence double-zeta basis with polarization functions. The Journal
  of chemical physics  \textbf{104}(20),  8007--8015 (1996)

\bibitem{frisch2009gaussian}
Frisch, M., Trucks, G., Schlegel, H., Scuseria, G., Robb, M., Cheeseman, J.,
  Scalmani, G., Barone, V., Mennucci, B., Petersson, G., Nakatsuji, H.,
  Caricato, M., Li, X., Hratchian, H.P., Izmaylov, A.F., Bloino, J., Zheng, G.,
  Sonnenberg, J.L., Hada, M., Ehara, M., Toyota, K., Fukuda, R., Hasegawa, J.,
  Ishida, M., Nakajima, T., Honda, Y., Kitao, O., Nakai, H., Vreven, T.,
  Montgomery, J.A.J., Peralta, J.E., Ogliaro, F., Bearpark, M., Heyd, J.J.,
  Brothers, E., Kudin, K.N., Staroverov, V.N., Kobayashi, R., Normand, J.,
  Raghavachari, K., Rendell, A., Burant, J.C., Iyengar, S.S., Tomasi, J., Cosi,
  M., Rega, N., Milla, J.M., Klene, M., Knox, J.E., Cross, J.B., Bakken, V.,
  Adamo, C., Jaramillo, J., Gomperts, R., Stratmann, R.E., Yazyev, O., Austin,
  A.J., Cammi, R., Pomelli, C., Ochterski, J.W., Martin, R.L., Morokuma, K.,
  Zakrzewski, V.G., Voth, G.A., Salvador, P., Dannenberg, J.J., Dapprich, S.,
  Daniels, A.D., Farkas, O., Foresman, J.B., Ortiz, J.V., Cioslowski, J., Fox:
  {Gaussian 09, Revision A. 02, 2009, Gaussian}. Inc., Wallingford CT  (2009)

\bibitem{hanwell2012avogadro}
Hanwell, M.D., Curtis, D.E., Lonie, D.C., Vandermeersch, T., Zurek, E.,
  Hutchison, G.R.: Avogadro: an advanced semantic chemical editor,
  visualization, and analysis platform. Journal of cheminformatics
  \textbf{4}(1),  1--17 (2012)

\end{thebibliography}

\end{document}